\documentclass[10pt]{article} 

\usepackage{aaspp4} 
\usepackage{psfig}

\slugcomment{\hbox to 6.5in{LBNL-42230 \hfil           \hfil
\hbox{ Presented at the January 1998 Meeting of the AAS}}}

\begin{document}
\tightenlines



\title{Cosmology from Type Ia Supernovae}
\author{
S.~Perlmutter, G.~Aldering,  S.~Deustua,
S.~Fabbro\altaffilmark{1}, G.~Goldhaber, D.~E.~Groom,
A.~G.~Kim\altaffilmark{2}, M.~Y.~Kim, R.A.~Knop, P.~Nugent,
C.~R.~Pennypacker} \vspace{0.025in}

\affil{\footnotesize Lawrence Berkeley National Laboratory and Center for
Particle Astrophysics, U.C. Berkeley}
\vspace{-0.1in} 

\author{A.~Goobar}
\vspace{-0.02in} \affil{\footnotesize University of Stockholm}
\vspace{-0.1in} 

\author{R.~Pain}
\vspace{-0.04in}\affil{$^1${\footnotesize LPNHE, CNRS-IN2P3 \& University of Paris VI \& VII}}
\vspace{-0.1in} 

\author{I.~M.~Hook, C.~Lidman}
\vspace{-0.02in} \affil{\footnotesize European Southern Observatory}
\vspace{-0.1in} 

\author{R.~S.~Ellis, M. Irwin, R.~G.~McMahon}
\affil{\footnotesize Institute of Astronomy, Cambridge}
\vspace{-0.1in} 

\author{P.~Ruiz-Lapuente}
\vspace{-0.02in}\affil{\footnotesize Department of Astronomy, University of Barcelona}
\vspace{-0.1in} 

\author{N.~Walton}
\vspace{-0.02in}\affil{\footnotesize Isaac Newton Group, La Palma}
\vspace{-0.1in} 

\author{B.~Schaefer}
\affil{\footnotesize Yale University}
\vspace{-0.1in} 

\author{B.~J.~Boyle}
\affil{\footnotesize Anglo-Australian Observatory}
\vspace{-0.1in} 

\author{A.~V.~Filippenko, T.~Matheson}
\affil{\footnotesize Department of Astronomy, U.C. Berkeley}
\vspace{-0.1in} 

\author{A.~S.~Fruchter, N.~Panagia}
\affil{\footnotesize Space Telescope Science Institute}
\vspace{-0.1in} 

\author{H.~J.~M.~Newberg}
\affil{\footnotesize Fermi National Laboratory}
\vspace{-0.1in} 

\author{W.~J.~Couch}
\affil{\footnotesize University of New South Wales}

\vspace{0.03in}
\author{(The Supernova Cosmology Project)}
\vspace{0.002in}


\begin{abstract}
\vspace{-0.2in} 

{\it This Lawrence Berkeley National Laboratory reprint is a reduction
of a poster presentation from the Cosmology Display Session \#85 on 
9 January 1998 at the American Astronomical Society meeting in 
Washington D.C.  It is also available on the World Wide Web at} \\
http://www-supernova.LBL.gov/  $\;\;\;\;\;\;$
{\it This work has also been referenced in the literature 
by the pre-meeting abstract
citation: Perlmutter et al., B.A.A.S., volume 29, page 1351 (1997).}
\vspace{0.08in}

This presentation reports on first evidence for a 
low-mass-density/positive-cosmological-constant universe that will
expand forever, based on observations of a set of 40 high-redshift 
supernovae.  The experimental strategy, data sets, and analysis techniques
are described.
More extensive analyses of these results with some additional 
methods and data
are presented in the more recent LBNL report \#41801 (Perlmutter et al., 1998; 
Ap.J., in press), astro-ph/9812133.

\end{abstract}

\psfig{file=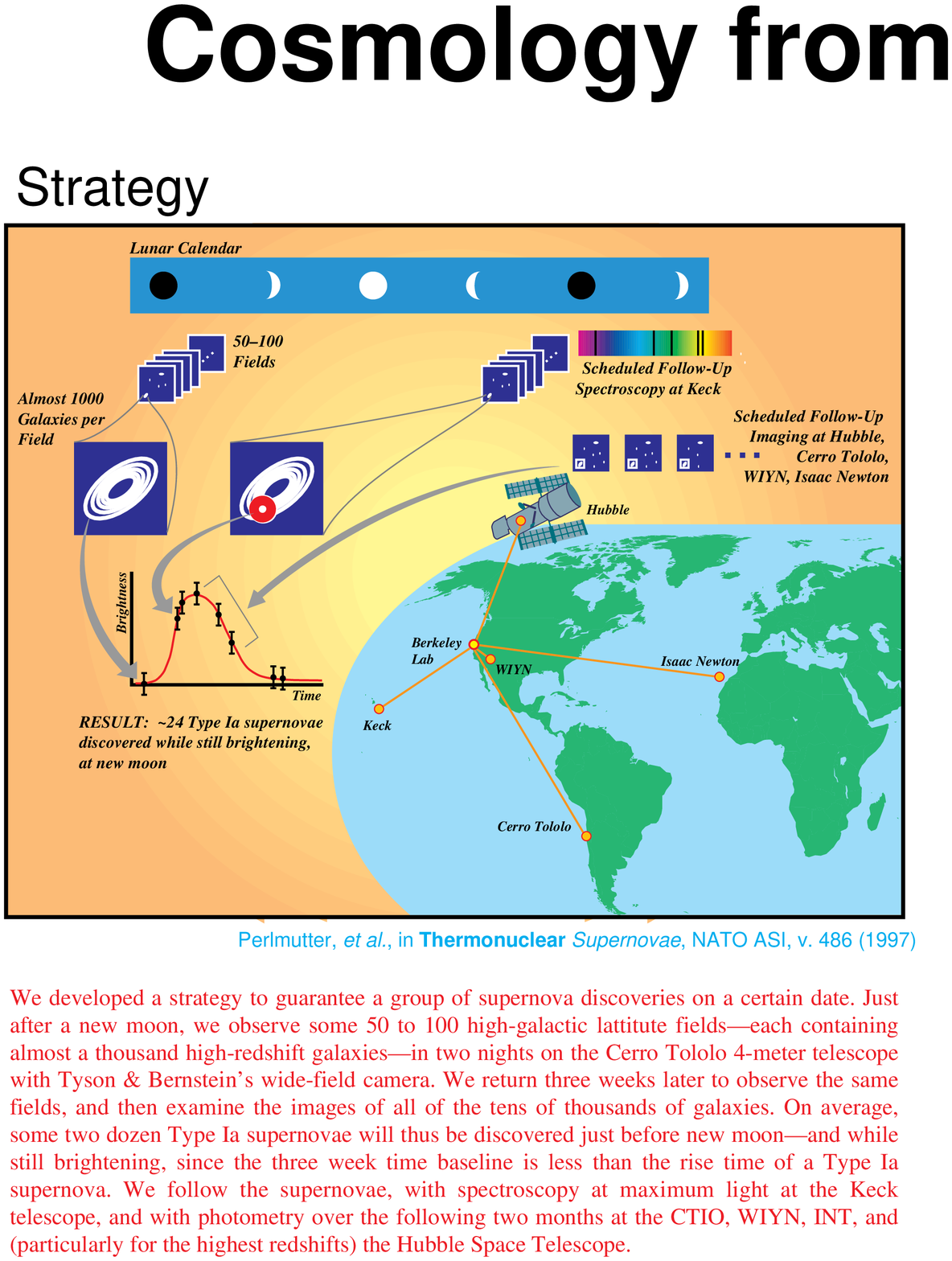,height=9.in}

\psfig{file=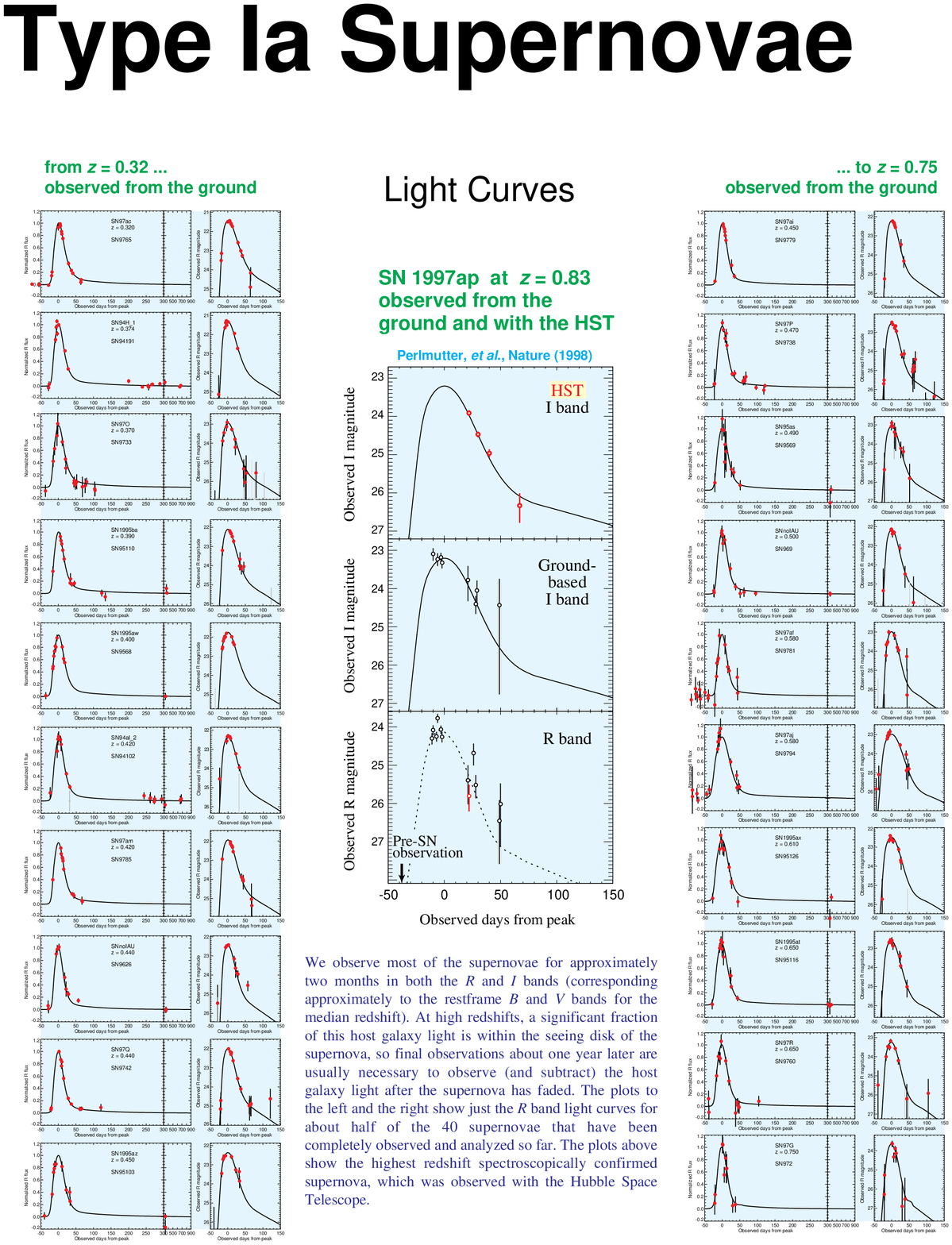,height=9.in}

\psfig{file=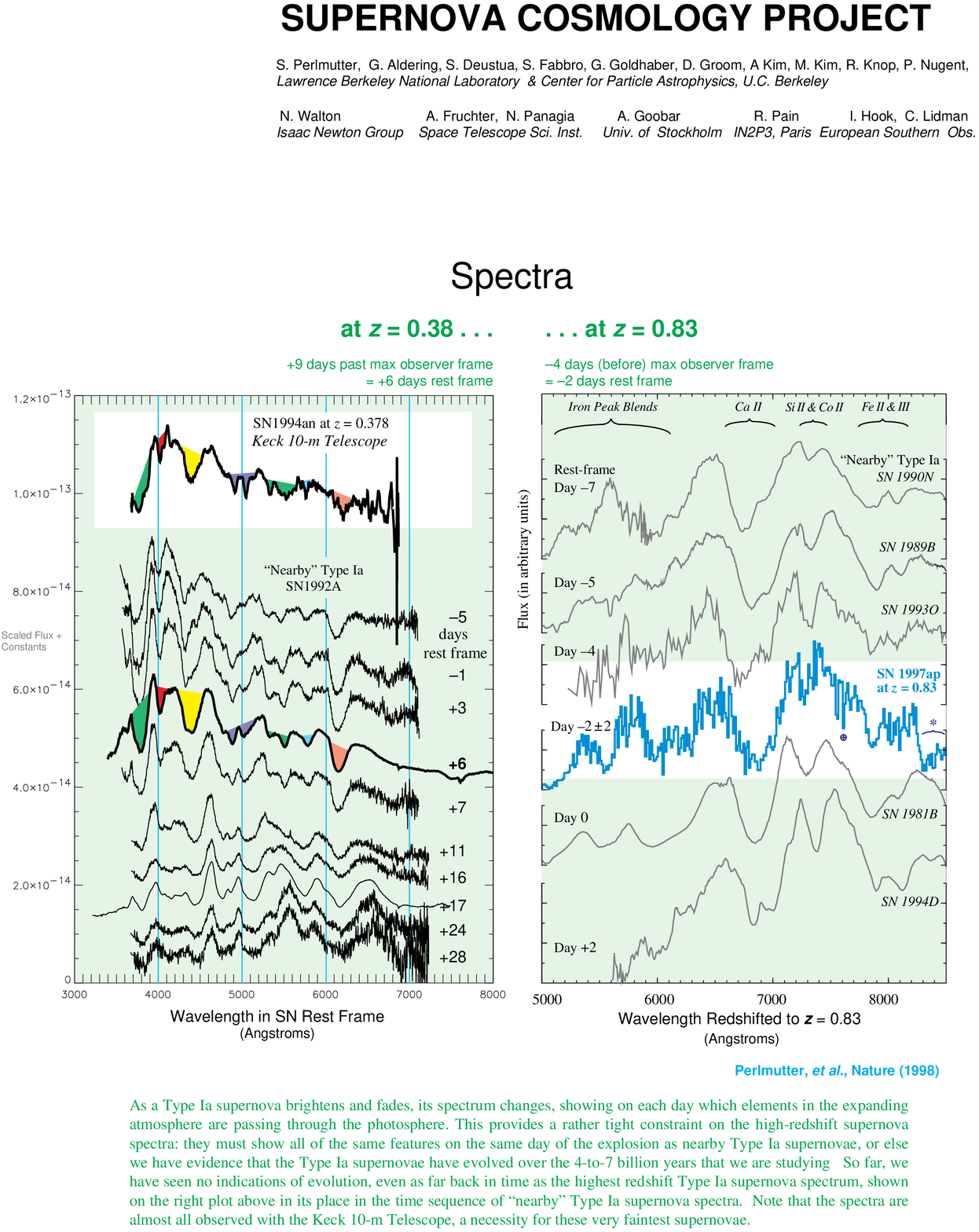,height=9.in}

\psfig{file=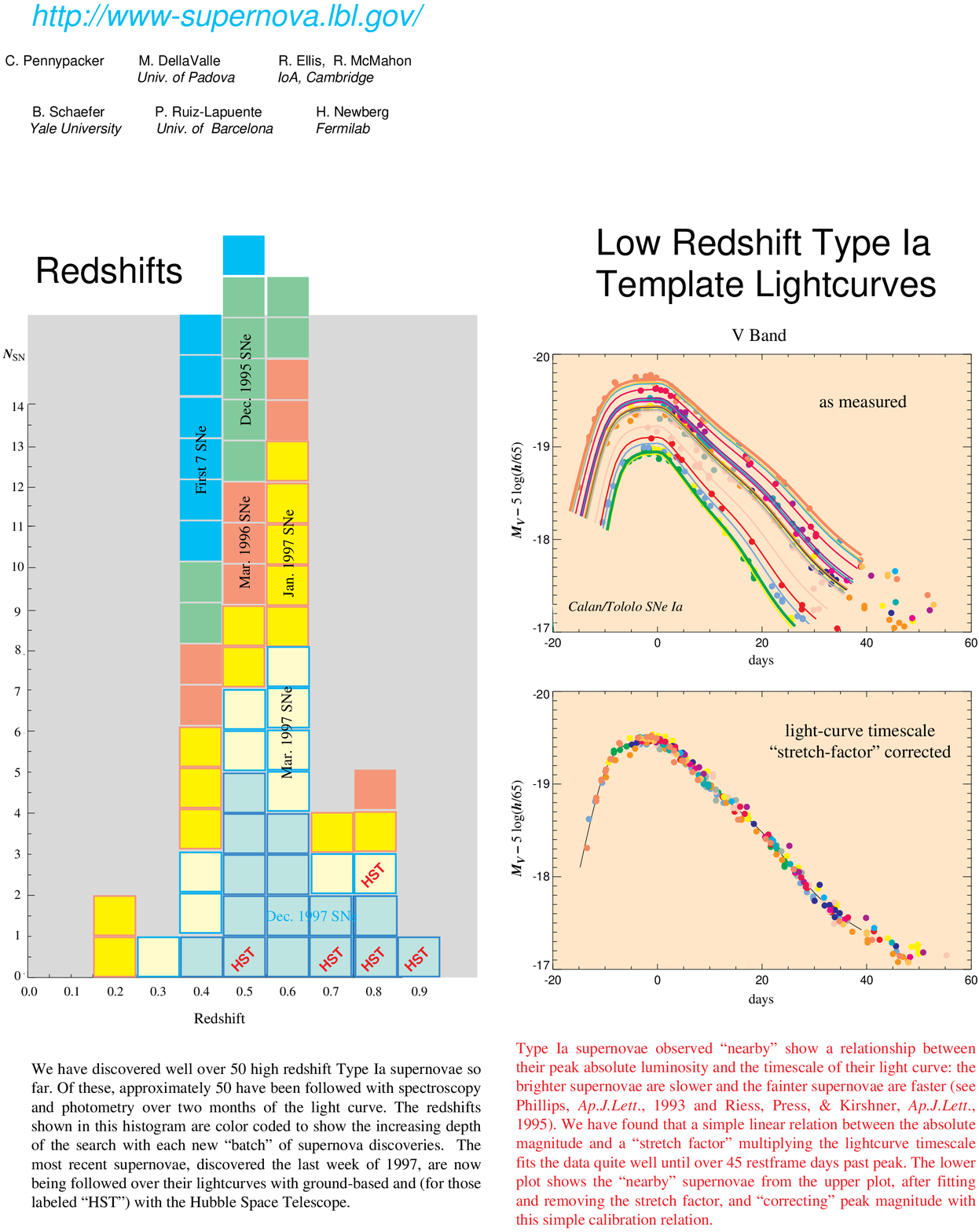,height=9.in}

\psfig{file=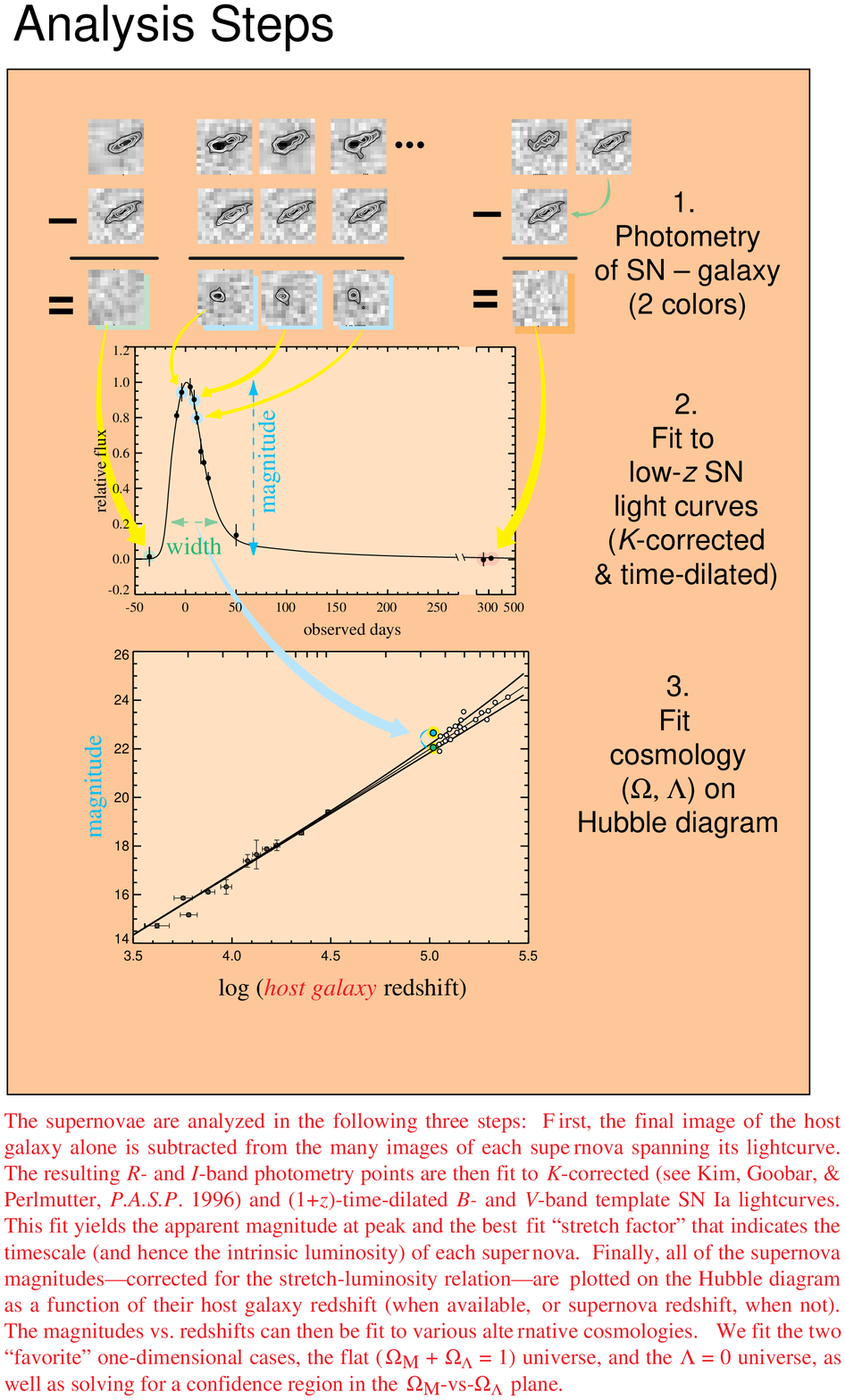,height=9.in}

\psfig{file=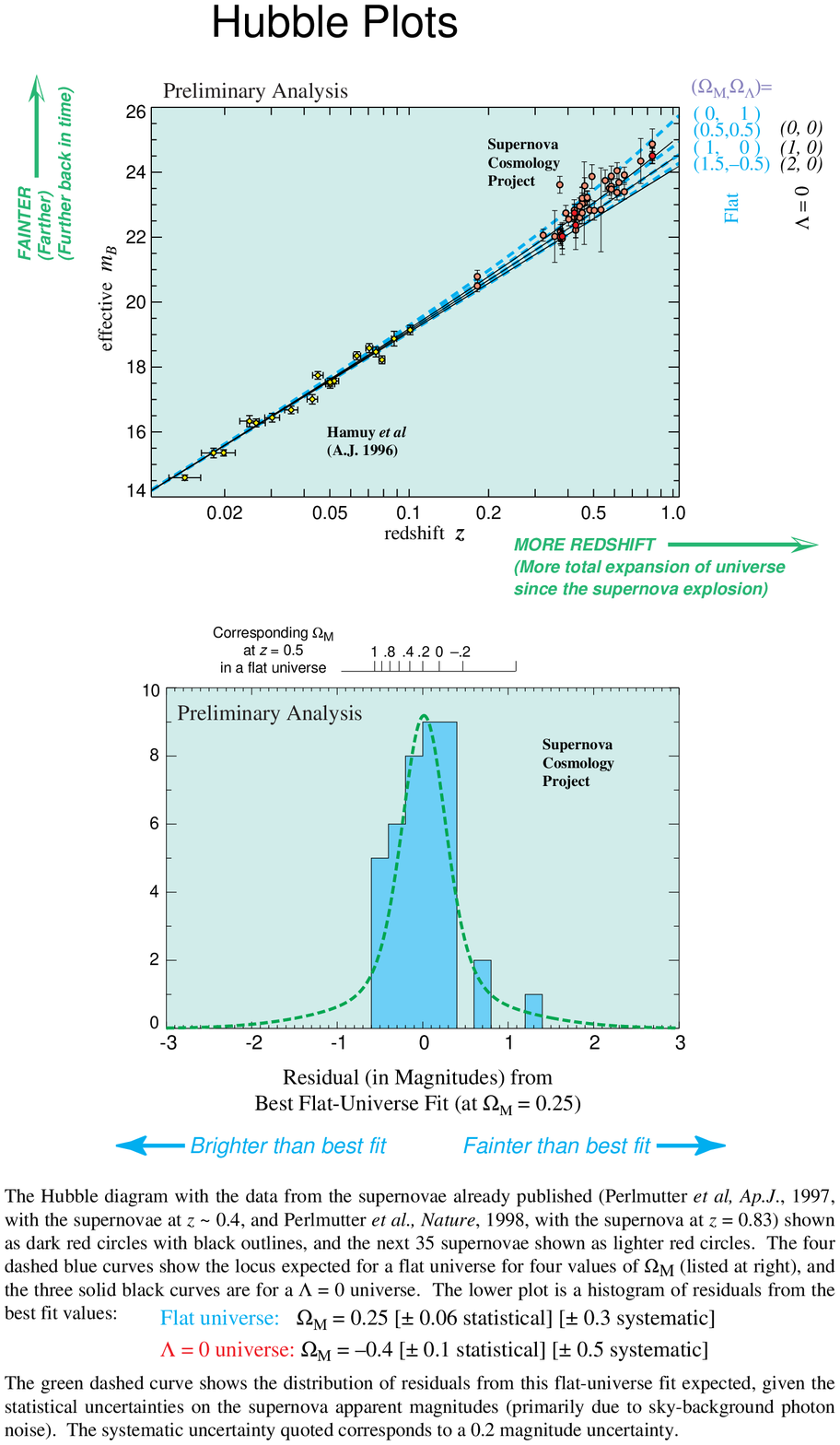,height=9.in}

\psfig{file=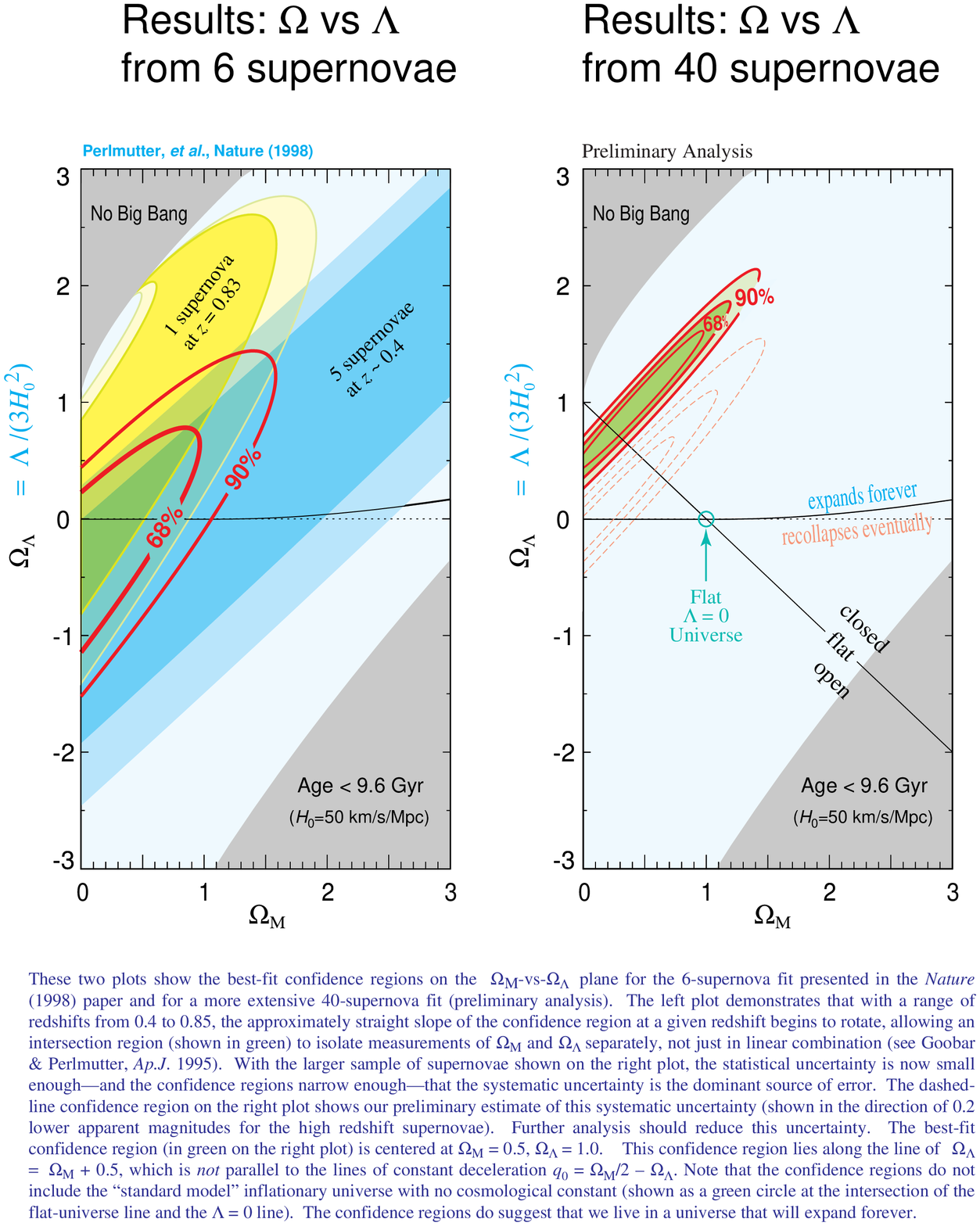,height=9.in}

\psfig{file=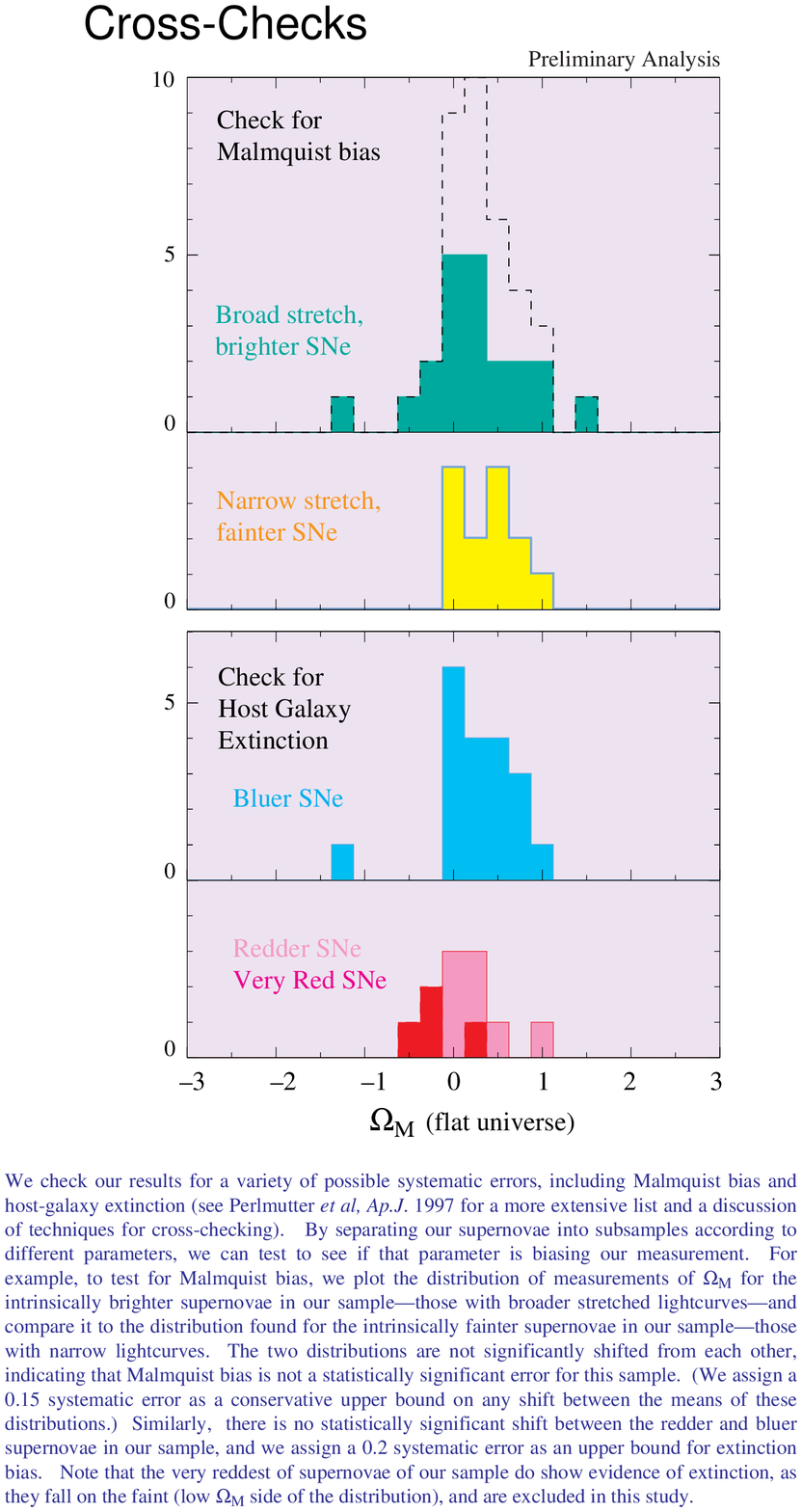,height=9.in}

\end{document}